\documentclass{article}
\usepackage{graphicx}
\usepackage[fleqn]{amsmath}
\usepackage{newtxtext}
\usepackage[varg]{newtxmath}

\usepackage{xurl}
\usepackage{circledsteps}

\title{A RAG Method for Source Code Inquiry Tailored to Long-Context LLMs}
\author{Toshihiro Kamiya\\Institute of Science and Engineering, Shimane University}

\begin{document}
\maketitle

\section{Introduction}

The development of text generation AI (large language models; LLMs) has been remarkable. LLMs have started to be used in software development as well, and AI assistant tools such as GitHub Copilot have emerged. However, there are still challenges with LLMs.

One of the issues frequently pointed out with LLMs is the context length limitation. The context length limitation refers to the upper bound on the length of context that an LLM can consider when generating text. For example, the LLM called \verb|gpt-4-32k| from OpenAI has a context length limitation of 32k tokens (approximately 60k characters in English). In user interfaces like ChatGPT's chat interface, it is common to implement an error when the input text exceeds this limit.

While LLMs that support long context input have emerged, even with such LLMs, according to research by Levy et al. \cite{Levy2024}, the inference performance of LLMs decreases as the input text becomes longer, leading to the so-called \textbf{needle in a haystack} problem.

Source code for software can easily exceed tens of thousands of lines, so inquiries that include source code may exceed the context length limitation, making it difficult to obtain high-quality answers.
In current software development, extensive reuse is practiced. Even if the product being developed has a small or compact source code, it is common for the code to expand by dozens of times when including the source code of the reused libraries, including frameworks.
Therefore, the context length limitation of LLMs becomes an issue when making inquiries such as identifying the cause of a bug or investigating the implementation of a specific feature.

One method to mitigate the context length limitation of LLMs is RAG (Retrieval-Augmented Generation) \cite{Lewis2021}.
In RAG, documents relevant to the inquiry in the prompt are retrieved and filtered from a database, web search, or some other method, and these documents are added to the original prompt as input to the LLM, allowing for answers based on the content of the documents.
For example, if the prompt is "Tell me about the habits of cats," the Wikipedia article "Cat" can be searched and its text added to the prompt to obtain a more detailed answer.
For inquiries about a software product, the source code of the product itself or the products being reused can be used as the documents to be added to the original prompt in RAG.

Another method to mitigate the context length limitation of LLMs is to use LLMs with larger contexts from the beginning. However, for the Transformer-based LLMs that are widely used today, the computational cost is proportional to the square of the context length, so increasing the context length directly leads to an increase in computational cost.

In this research, we propose a RAG method for inquiries about the source code of software products. The proposed method aims to mitigate and solve the needle in a haystack problem by obtaining accurate answers without referring to the entire source code, by executing the product to obtain an execution trace (log of called functions), extracting the call tree and source code of the called functions from the execution trace, and inputting them to the LLM as documents for RAG.

\section{Related Research}

As research on applying LLMs to software development, a study \cite{Jin2023} on using LLMs for bug localization proposes a method where error messages from failed test cases are input to the LLM to identify the cause, and further, to identify the location of the bug.

Tools have been proposed that incorporate LLMs into IDEs to assist developers in understanding source code and APIs \cite{Nam2024}. The goal is to improve the efficiency of developers' work by answering their questions and automatically displaying summaries of the code, without interrupting their concentration.

The literature \cite{Xia2024} points out that existing benchmarks may not be able to accurately evaluate the performance of LLMs because the training data includes answer examples. This issue is also partially encountered in the experiment described in Section \ref{se:exp1}.

A blog post\footnote{Using GitHub Copilot in your IDE: Tips, tricks, and best practices, \url{https://github.blog/2024-03-25-how-to-use-github-copilot-in-your-ide-tips-tricks-and-best-practices/}} for the GitHub Copilot software development AI assistant tool states that to obtain appropriate responses from the AI, you should open files related to the task you are currently working on, in order to provide the LLM with the appropriate context.
The method proposed in this research automatically identifies the source code that should be shown to the LLM by dynamically analyzing the target product.

\section{Approach}

The proposed method takes a user inquiry and the execution trace of the software as input, and identifies the corresponding source code through the following steps, which is then input to the LLM.

\noindent
\textbf{Step 1.} The user inputs an inquiry about a software product and the trace (log of called functions) collected when executing the feature related to that inquiry.
The execution trace is obtained using the trace collection tool \verb|rapt| developed by the author. This tool executes the target Python script and collects a log of the called functions.
The location information (source file and line number) of the functions in the source code is also collected.
For example, in the experiment described in Section \ref{se:exp1}, an inquiry is made about the feature of formatting and displaying CSV files in the \verb|rich-cli| command-line interface (CLI) tool, and the execution trace is collected by displaying a CSV file.

\noindent
\textbf{Step 2.} Analyze the execution trace and identify the executed functions and methods, as well as their call relationships.

\noindent
\textbf{Step 3.} Identify the corresponding source code file and the location of the function within it from the names of the called functions.

\noindent
\textbf{Step 4.} Create a call graph (a graph representing the call relationships between functions) from the call relationships, and further generate a call tree (tree structure) by removing loops and other constructs.

\noindent
\textbf{Step 5.} Input the prompt, which is the inquiry text appended with the call tree and the source code of the functions within the call tree, to the LLM and output the response.

In this case, the source code of the functions is arranged in the order they appear in the call tree. In other words, if a function \verb|f| calls another function \verb|g|, the source code of \verb|g| is placed after the source code of \verb|f|. This ordering is intended to make it easier for the LLM to follow the execution flow by showing the lines of source code in the order they are executed. The experiment described in Section \ref{se:exp1} evaluates the effect of this ordering.

\subsection{Practical Considerations}

In the implementation of the proposed method, the following processing is performed to represent the call tree in a compact form:

\noindent
(1) When a function calls the same function multiple times, the called function is represented as a single node. Also, when a function is called from different functions, the called function nodes are represented separately (not merged).

As a result, such a call tree becomes a tree obtained by removing recursion (i.e., cycles in the graph) and merging (i.e., parts of the graph where nodes have multiple incoming edges) from the call graph represented as a directed graph. By representing it as a tree structure, it is expected to become easier to determine the order in which to append the source code to the prompt.

\noindent
(2) When there is a recursive call, the hierarchy up to the node where the first recursive call occurs is represented, and nodes for deeper recursive calls are omitted. Recursive calls are sometimes used for iteration, and this avoids generating an excessively large call tree in such cases.

\noindent
(3) Other considerations and limitations

The execution trace collection tool \verb|rapt| only records the calls of functions within the modules specified by the user. Calls to built-in functions (e.g., \verb|print|) or functions in the standard library are not recorded.

\verb|rapt| adopts a method of wrapping functions and logging their calls during the execution of the target product. However, due to the limitations of this implementation method, it is not possible to record calls to functions of dynamically (lazily) loaded modules or functions that are not in the global scope (such as lambdas).

In general, programs perform module imports and initialization during startup.
Since there may be cases where such processing needs to be excluded from the analysis, a branch pruning function for the call tree was implemented.
To perform branch pruning, first, an execution trace is obtained by executing the program without running its functionality (by immediately exiting after program start).
The functions included in such a call tree are considered to be related to startup, and if the same node (same function call) is present in the leaves of the call tree being analyzed, it is removed.

\section{Experiments}

In this section, we evaluate the proposed method by applying it to an open-source product. We prepared specific inquiries that could occur during software development and criteria for evaluating the quality of responses to those inquiries (described later). We create variants by changing the content of the prompt (presence or order of the call tree and function source code) and analyze whether the proposed method contributes to the quality of the response by making inquiries with these variants and evaluating the quality of the responses (evaluation criteria described later).

\subsection{Target Product}

For the experiment, we selected the OSS command-line tool \verb|rich-cli|\footnote{rich-cli \url{https://github.com/Textualize/rich-cli}} as the target. The reasons for choosing this tool are that it is feature-rich and has an appropriate scale (approximately 220k lines) for the experiment, and because it is a CLI tool, it is easy to maintain consistent conditions when running repeatedly (high reproducibility).
The \verb|rich-cli| tool has the ability to take files in formats such as CSV, Markdown, and ReStructuredText as input, and format and display them in the terminal with syntax highlighting, among other features.

Table \ref{tab:targetproduct} shows the line counts (measured using the \verb|cloc|\footnote{Cloc \url{https://github.com/AlDanial/cloc}} tool) of Python source files for the target product and the libraries (packages) it directly and indirectly depends on. The target product \verb|rich-cli| itself is only 900 lines, as it is a tool that allows the functionality of libraries such as \verb|rich| and \verb|rich-rst| to be accessed from the command line. However, including the dependent libraries, the total is approximately 220k lines.

\begin{table}[htb]
    \centering
    \caption{Packages that make up the target product} \label{tab:targetproduct}
    \begin{tabular}{llr}
        \hline
        Package & Version & Python Lines \\ \hline
        certifi & 2024.2.2 & 63 \\
        charset-normalizer & 3.3.2 & 4,022 \\
        click & 8.1.7 & 5,659 \\
        docutils & 0.20.1 & 28,303 \\
        idna & 3.6 & 11,142 \\
        linkify-it-py & 2.0.3 & 2,032 \\
        markdown-it-py & 3.0.0 & 4,226 \\
        mdit-py-plugins & 0.4.0 & 2,440 \\
        mdurl & 0.1.2 & 342 \\
        Pygments & 2.17.2 & 94,240 \\
        requests & 2.31.0 & 2,904 \\
        rich & 13.7.1 & 19,638 \\
        rich-cli & 1.8.0 & 900 \\
        rich-rst & 1.2.0 & 569 \\
        textual & 0.54.0 & 33,438 \\
        typing\_extensions & 4.10.0 & 1,633 \\
        uc-micro-py & 1.0.3 & 14 \\
        urllib3 & 2.2.1 & 6,419 \\ \hline
        (Total) & ~ & 217,984 \\ \hline
    \end{tabular}
\end{table}

\subsection{LLMs Used in the Experiment}

Table \ref{tab:gais} shows the context length limits (in tokens) of the LLMs Gemini 1.5 Pro \cite{Reid2024}, Claude 3 Sonnet, and ChatGPT-4 used in the experiment.
As described in Section \ref{se:exp2} below, the length of the generated prompts can reach up to around 87k tokens, so LLMs capable of handling relatively large contexts were selected.
In all cases, the responses were generated by pasting the prompts into a chat-style UI, without using an API.
Each LLM uses a different tokenizer (a function that splits text into tokens), so the token count for the same text will differ, making the context length limit only a rough guideline.

\begin{table}[htb]
    \centering
    \caption{LLMs used in the experiment} \label{tab:gais}
    \begin{tabular}{lp{0.1\linewidth}l}
        \hline
        LLM             & c.   & Service Used             \\ \hline
        Gemini 1.5 Pro  & 1M   & Accessed from Google AI Studio \\
        Claude 3 Sonnet & 200K & Accessed from poe.com          \\
        ChatGPT-4       & 128K & Accessed from poe.com          \\ \hline
    \end{tabular}\\
    \footnotesize{The c. column shows the context length limit (in tokens).}
\end{table}

\subsection{Prompt Variants}

\begin{figure}[htb]
    \begin{center}
        \includegraphics[width=0.7\linewidth]{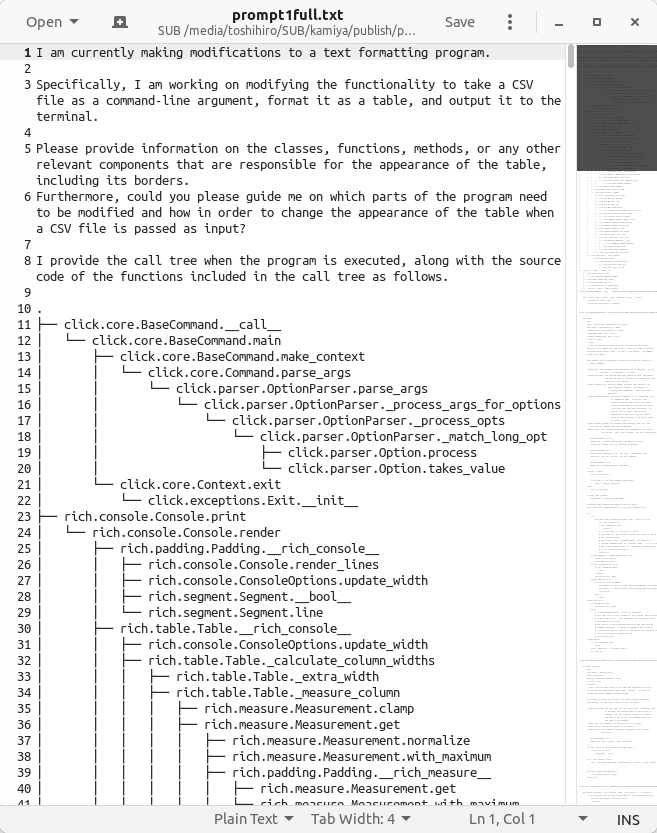}
    \end{center}
    \caption{Example prompt} \label{fig:prompt1full}
\end{figure}

The proposed method includes the call tree and source code in the prompt. To experimentally evaluate the effect of including these in the prompt, we create variants by removing or reordering them from the prompt generated by the proposed method and evaluate the quality of the responses.
Table \ref{tab:promptv} shows the variants of the prompt used in the experiment. Figure \ref{fig:prompt1full} shows an example of the prompt used in the following Section \ref{se:exp1}.

\begin{table}[htb]
    \centering
    \caption{Prompt Variants} \label{tab:promptv}
    \begin{tabular}{p{0.08\linewidth}p{0.9\linewidth}}
        \hline
        Variant & Description \\ \hline
        full & The prompt generated by the proposed method itself, including the inquiry text, call tree, and source code of the functions within the call tree in the order they appear in the call tree. If the same function appears multiple times in the call tree, its source code is included multiple times in the prompt. \\ \hline
        A    & Same as full, but the source code within the call tree is sorted by function name (including module name). If the same function appears multiple times in the call tree, its source code is included only once in the prompt without duplication. \\ \hline
        C    & Excludes the call tree from full. The source code of the functions is included in the order they appear in the call tree.                                         \\ \hline
        CA   & Excludes the call tree from full. The source code of the functions is sorted and duplicates are removed.                                               \\ \hline
        T    & Excludes the source code from full.                                                                                                   \\ \hline
    \end{tabular}
\end{table}

\subsection{Experiment 1} \label{se:exp1}

We create prompts following the proposed method for specific inquiries that could arise during software development, input them to LLMs, and evaluate their responses according to the evaluation criteria.
In this experiment, the response generation is performed only once for each prompt variant and each LLM. With the chat-style UI of the LLM services, a random number is used, so the content of the response differs each time it is generated. In that sense as well, the evaluation of the responses is not absolute, but should be noted as an assessment of the overall trend.

\subsubsection{Prompt 1}

The tool has a feature to format and display CSV files in the terminal. This feature uses line characters to create the appearance of a table.
The inquiry asks what function determines this appearance (such as the type of line or right alignment) and how to change the appearance.
The command line of the tool does not provide a way to change the format, so the source code needs to be modified.
The correct answer for the location to be modified was determined by considering the ability to confirm that the change would modify the functionality of the product, that no modifications other than the indicated location are necessary, and that the impact on other features of the product should be minimized as much as possible.

The author visually inspected each response and evaluated it on a 4-point scale according to the following criteria. Items in the response that were deemed to be incorrect explanations or hallucinations were deducted points.

\noindent
\Circled{1} The name of the class or function that implements the feature can be output (1 point). If multiple instances are mentioned and the correct one is included, 0.5 points.

\noindent
\Circled{2} In addition to lines, elements such as color and padding that affect the appearance of the table are explained (1 point).

\noindent
\Circled{3} The content of the change (the modified code or how to modify it) can be output (1 point). If multiple instances are mentioned and the correct one is included, 0.5 points.

\noindent
\Circled{4} The location to be modified can be output by function name (1 point). If multiple instances are mentioned and the correct one is included, 0.5 points.

The results for Prompt 1 are shown in Table \ref{tab:p1results}. The evaluation scores for the responses obtained from all LLMs and all variants were high.
However, for the T variant, which included only the inquiry text and the call tree where function names are nodes, some responses included the name of a global variable.
This could be due to hallucination or the possibility that the LLM's training data included information about the \verb|rich| package, causing the response to be generated based on the learned information rather than the content of the prompt (\textbf{suspected data leakage}).

\begin{table}[htb]
    \centering
    \caption{Evaluation scores for Prompt 1 responses} \label{tab:p1results}
    \begin{tabular}{llllll}
        \hline
        LLM             & full & A   & C   & CA  & T    \\ \hline
        Gemini 1.5 Pro  & 4.0  & 4.0 & 4.0 & 3.0 & 4.0* \\
        Claude 3 Sonnet & 4.0  & 4.0 & 4.0 & 4.0 & 4.0  \\
        ChatGPT-4       & 3.0  & 4.0 & 4.0 & 3.0 & 4.0* \\ \hline
    \end{tabular}\\
    \footnotesize{The numbers marked with * indicate that the name of a global variable not included in the call tree was included in the response.}
\end{table}

\subsubsection{Prompt 2}

The tool has a feature to format and display Markdown files containing tables in the terminal. Similar to Prompt 1, we inquired about how to change the appearance.

Since Markdown has many formatting options besides tables, such as lists and quotes, the parsing performed is more complex compared to processing CSV files. As a result, there are more classes and methods involved in the processing, making it more difficult than Prompt 1.
The same evaluation criteria as Prompt 1 were used.

The results for Prompt 2 are shown in Table \ref{tab:p2results}. For all LLMs, the maximum evaluation score was obtained from the responses generated by the full or A variants.
Additionally, as with Prompt 1, the responses for the T variant included suspected data leakage.

\begin{table}[htb]
    \centering
    \caption{Evaluation scores for Prompt 2 responses} \label{tab:p2results}
    \begin{tabular}{llllll}
        \hline
        LLM             & full & A   & C   & CA  & T    \\ \hline
        Gemini 1.5 Pro  & 4.0  & 4.0 & 4.0 & 2.0 & 3.0* \\
        Claude 3 Sonnet & 2.0  & 3.5 & 1.5 & 2.5 & 2.5  \\
        ChatGPT-4       & 4.0  & 3.0 & 3.0 & 3.0 & 3.5* \\ \hline
    \end{tabular}\\
    \footnotesize{The numbers marked with * indicate that the name of a global variable not included in the call tree was included in the response.}
\end{table}

\subsubsection{Prompt 3} \label{se:prompt3}

We compared the feature used in Prompt 1 to format and display CSV files in the terminal and the feature used in Prompt 2 to format and display Markdown files containing tables in the terminal.
Specifically, we inquired about the differences and similarities in implementation, differences in control flow and data structures, and differences in table-related functionality.
Prompt 3 was the longest in the experiment and serves as a benchmark for testing the scalability of the proposed method. The following evaluation criteria were used:

\noindent
\Circled{1} Differences and similarities in implementation are explained (1 point).

\noindent
\Circled{2} Names of important functions or methods that illustrate differences in control flow are provided. 1 point if both are correct. 0.5 points if one is correct.

\noindent
\Circled{3} Differences in the data structures used are explained by class names. 1 point if both are correct. 0.5 points if one is correct.

\noindent
\Circled{4} Differences in table-related functionality between the two (e.g., right alignment) are explained (1 point). General Markdown features (such as links) are excluded. If specific examples like right alignment are not provided and only terms like "styles" are used, 0.5 points.

Since Prompt 3 requires comparing two executions, the content of the prompt was arranged in the following order: the inquiry text, the call tree of the first execution, the sequence of source code of functions included in the call tree of the first execution, the call tree of the second execution, and the sequence of source code of functions included in the call tree of the second execution.
However, for the CA variant prompt, the content was the inquiry text and the source code of functions included in either execution call tree, sorted by function name.

The results for Prompt 3 are shown in Table \ref{tab:p3results}. The low evaluation score for the full variant of ChatGPT-4 is notable.
This is mere speculation, but since the full variant of Prompt 3 is also the longest prompt in this experiment (as described in Section \ref{se:exp2}), it may have approached the LLM's context length limit, resulting in a degradation of response quality (\textbf{suspected context length limit issue}).

\begin{table}[htb]
    \centering
    \caption{Evaluation scores for Prompt 3 responses} \label{tab:p3results}
    \begin{tabular}{llllll}
        \hline
        LLM             & full & A   & C   & CA  & T   \\ \hline
        Gemini 1.5 Pro  & 3.5  & 4.0 & 2.0 & 3.0 & 1.5 \\
        Claude 3 Sonnet & 2.0  & 2.0 & 1.5 & 2.0 & 3.0 \\
        ChatGPT-4       & 1.0  & 3.5 & 2.5 & 1.0 & 3.0 \\ \hline
    \end{tabular}
\end{table}

\subsubsection{Prompt 4}

The tool has a command line option \verb|--emoji| that converts emoji codes (\verb|:sparkle:|, etc.) to actual emoji when outputting text. This option works when the text is provided as a command line argument, but not when it is provided from a file. We inquired about the cause and how to fix it.

Since the implementation differs between text provided as a command line argument and text provided from a file, the emoji processing needs to be added to the file processing part. Branching based on the presence of the option is also necessary. Additionally, to maintain consistency in the tool's functionality, modifications that reuse existing processing are preferable, so reuse was added as an evaluation item. Prompt 4 requires making a design decision on where to add the necessary processing within the product, making it more difficult compared to the previous prompts.

The evaluation criteria were as follows:

\noindent
\Circled{1} The cause can be output (1 point). If multiple causes are mentioned and the correct one is included, 0.5 points.

\noindent
\Circled{2} The content of the change (the modified code or how to modify it) can be output (1 point). If multiple instances are mentioned and the correct one is included, 0.5 points.

\noindent
\Circled{3} The location to be modified can be output by function name (1 point). If multiple instances are mentioned and the correct one is included, 0.5 points.

\noindent
\Circled{4} A modification plan that reuses existing functionality can be output (1 point). If the plan is to create new functionality, 0.5 points.

The results for Prompt 4 are shown in Table \ref{tab:p4results}. For all LLMs, the maximum evaluation score was obtained from the full variant (however, for ChatGPT-4, the evaluation score was 2.5 for all variants).

\begin{table}[htb]
    \centering
    \caption{Evaluation scores for Prompt 4 responses} \label{tab:p4results}
    \begin{tabular}{llllll}
        \hline
        LLM             & full & A   & C   & CA  & T   \\ \hline
        Gemini 1.5 Pro  & 4.0  & 2.5 & 2.0 & 3.0 & 1.0 \\
        Claude 3 Sonnet & 4.0  & 3.0 & 2.0 & 2.0 & 3.0 \\
        ChatGPT-4       & 2.5  & 2.5 & 2.5 & 2.5 & 2.5 \\ \hline
    \end{tabular}
\end{table}

\subsubsection{Analysis}

\begin{figure}[htb]
    \begin{center}
        \includegraphics[width=0.55\linewidth]{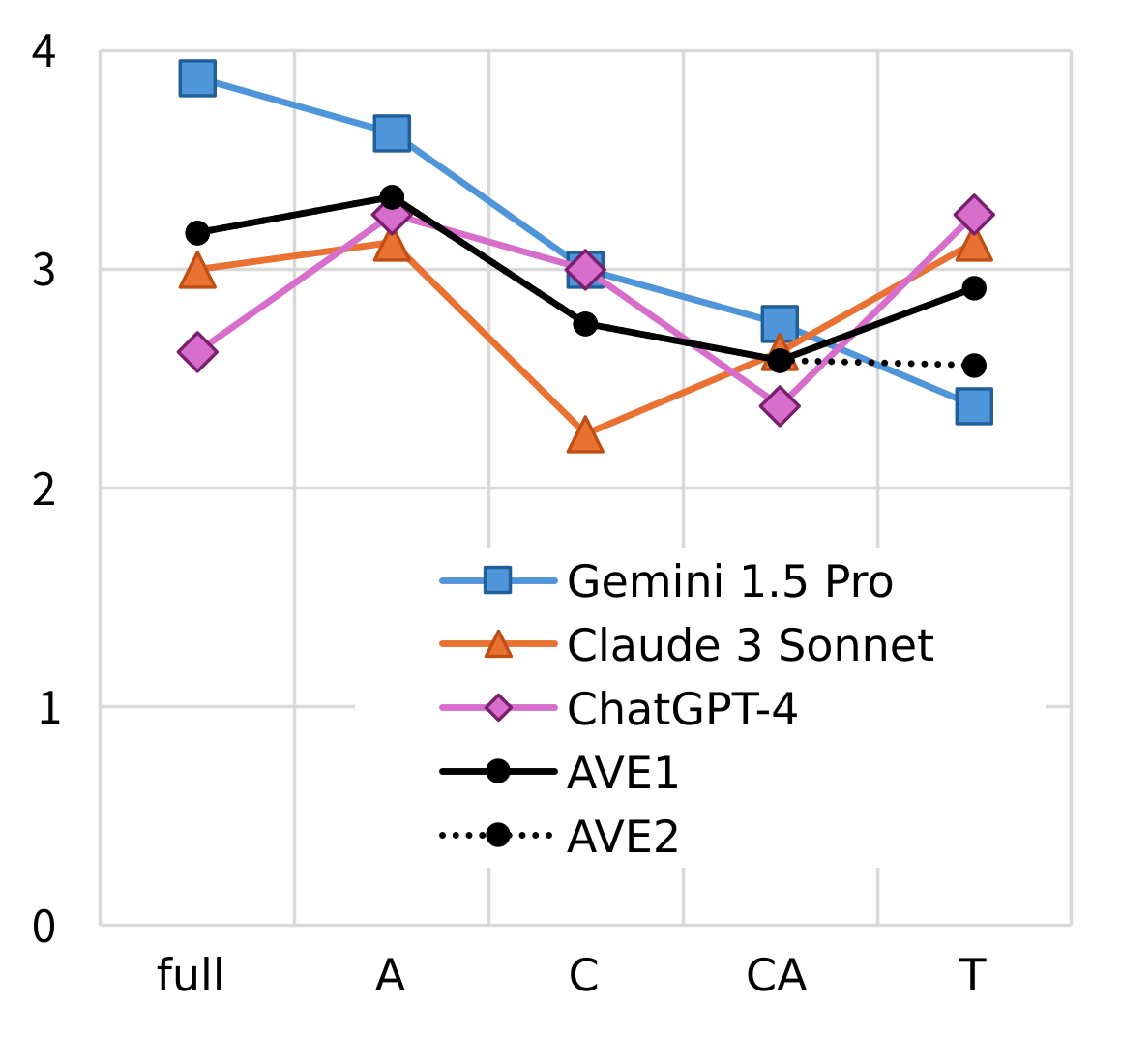} 
    \end{center}
    \caption{Trend of evaluation scores} \label{fig:overallresults}
\end{figure}

Figure \ref{fig:overallresults} shows a line graph representing the average of Prompts 1 to 4 for each variant (full, A, C, CA, T), with the horizontal axis representing the prompt variants and the vertical axis representing the aggregated evaluation scores.
The solid black line (AVE1) uses the raw experimental results, while the black dotted line (AVE2) excludes the suspected data leakage observed in Prompts 1 and 2.
For example, the black circle on the far left indicates that the average evaluation score for the full variant across all LLMs and all prompts (1 to 4) is 3.17.

The lines other than the solid black line show the scores aggregated and classified by LLM.
The drop in evaluation scores at full could be due to the influence of the context length limit, as observed in Prompt 3 of Experiment 1.

Since this experiment has a small sample size, statistical judgments cannot be made, but overall trends can be discussed.

First, looking at Gemini 1.5 Pro, which has the largest context, the evaluation score for the full variant is the highest, followed by A, C, CA, and T in that order.

Focusing on the presence or absence of source code, the evaluation score for the T variant, which does not include source code, does not reach the scores of the full or A variants, which include both source code and the call tree.
This suggests a trend that including the call tree and source code in the prompt contributes to the quality of the response.

Looking at the variants that include source code, i.e., full, A, C, and CA, the only variant that does not include the order in which functions are called is CA. The other variants, full, A, and C, include information about the order in which functions are called, either through the call tree or the ordering of the function source code.
The lower evaluation score for the CA variant suggests a trend that the order in which functions are called contributes to the quality of the response.

Further examining the full, A, and C variants, the C variant presents the order in which functions are called through the ordering of the source code, while full and A present the order in which functions are called through the call tree.
Excluding the suspected context length limit issue described in Prompt 3 of Experiment 1, there appears to be a trend that presenting the order through the call tree contributes to the quality of the response.

\subsection{Experiment 2} \label{se:exp2}

We evaluate the size of the prompts generated by the proposed method.

\noindent
(1) Examine how the prompt size compares to the LLM's context length limit.

Table \ref{tab:promptsize} shows the prompt lengths of Prompts 1 to 4 measured using the ChatGPT-4 tokenizer\footnote{The Tokenizer Playground https://huggingface.co/spaces/Xenova/the-tokenizer-playground was used.}. The maximum is the full variant of Prompt 3 with 87,950 tokens. Theoretically, this reaches nearly 70\% of the context length limit of the ChatGPT-4 used in this experiment. Additionally, when measured with the Gemini (Gemma) tokenizer, this prompt becomes 106,875 tokens, revealing a difference of around 20\% in token counts between LLMs.

\begin{table}[htb]
    \centering
    \caption{Prompt sizes (in ChatGPT-4 tokens)} \label{tab:promptsize}
    \begin{tabular}{lrrrrr}
        \hline
        Prompt  & full   & A      & C      & CA     & T     \\ \hline
        Prompt 1 & 32,250 & 19,949 & 22,079 & 18,768 & 1,279 \\
        Prompt 2 & 64,838 & 53,537 & 61,238 & 49,943 & 3,711 \\
        Prompt 3 & 87,950 & 73,338 & 83,198 & 50,528 & 4,876 \\
        Prompt 4 & 26,104 & 23,984 & 24,489 & 18,792 & 1,751 \\ \hline
    \end{tabular}
\end{table}

In terms of prompt size, the full variant is the largest for all prompts from 1 to 4. The C variant is smaller because it does not include the call graph. The A variant is smaller because it does not include duplicate function source code. The T variant is smaller because it does not include source files.

\noindent
(2) Examine how the prompt size compares to the source files of the target product.

Table \ref{tab:srcsize} shows the number of Python source files included in Prompts 1 to 4 and the total number of lines in those source files. For comparison, the number of lines in the full variant prompt is also included. Table \ref{tab:targetproduct} showed that the total line count for the target product was approximately 220k lines, so in comparison to inputting the entire source code of the target product into the LLM, the number of lines in the prompts is less than 1/20.

\begin{table}[htb]
    \centering
    \caption{Python source code referenced by the prompts} \label{tab:srcsize}
    \begin{tabular}{lrr|r}
        \hline
        Prompt  & File Count & File Lines & Full Variant Lines \\\hline
        Prompt 1 & 14         & 11,552       & 2,549 \\
        Prompt 2 & 53         & 18,799       & 7,079 \\
        Prompt 3 & 53         & 18,799       & 9,626 \\
        Prompt 4 & 18         & 13,757       & 2,737 \\ \hline
    \end{tabular}\\
    \footnotesize{Prompts 2 and 3 reference the same set of source files, \\hence the file count and file lines are the same.}
\end{table}

With the proposed method, the necessary source files are identified by executing the target product, eliminating the need for the user (developer) to manually select source files. Even if developers could select source files, the proposed method extracts the source code at the function level and appends it to the prompt, allowing for smaller prompts. For example, the full variant of Prompt 1 is 22\% (=2549/11552) in terms of line count.

\section{Summary and Future Prospects}

In this research, we proposed a RAG method for inquiries about source code. The proposed method extracts the call tree and the source code of the called functions from the execution trace of a software product and appends them to the prompt. This enables inquiries that require considering the product's design, such as investigating differences in functionality or determining where functionality should be implemented.

In the experiment, we used an open-source product of approximately 220k lines, including dependencies, as the target and evaluated the responses by inputting the created prompts into LLMs for specific inquiries that could arise during software development. The experimental results showed a trend of improved response quality when including the call tree and source code in the prompt. In particular, it was found that including the order in which functions are called in the prompt is important.
On the other hand, there were cases where the quality of the response degraded when the prompt size became large, depending on the LLM.

Future tasks include automating prompt generation to reduce manual effort by the user and establishing a method for creating prompts that can handle various software tasks. It will also be necessary to investigate more effective methods for addressing the context length limit of LLMs.

\end{document}